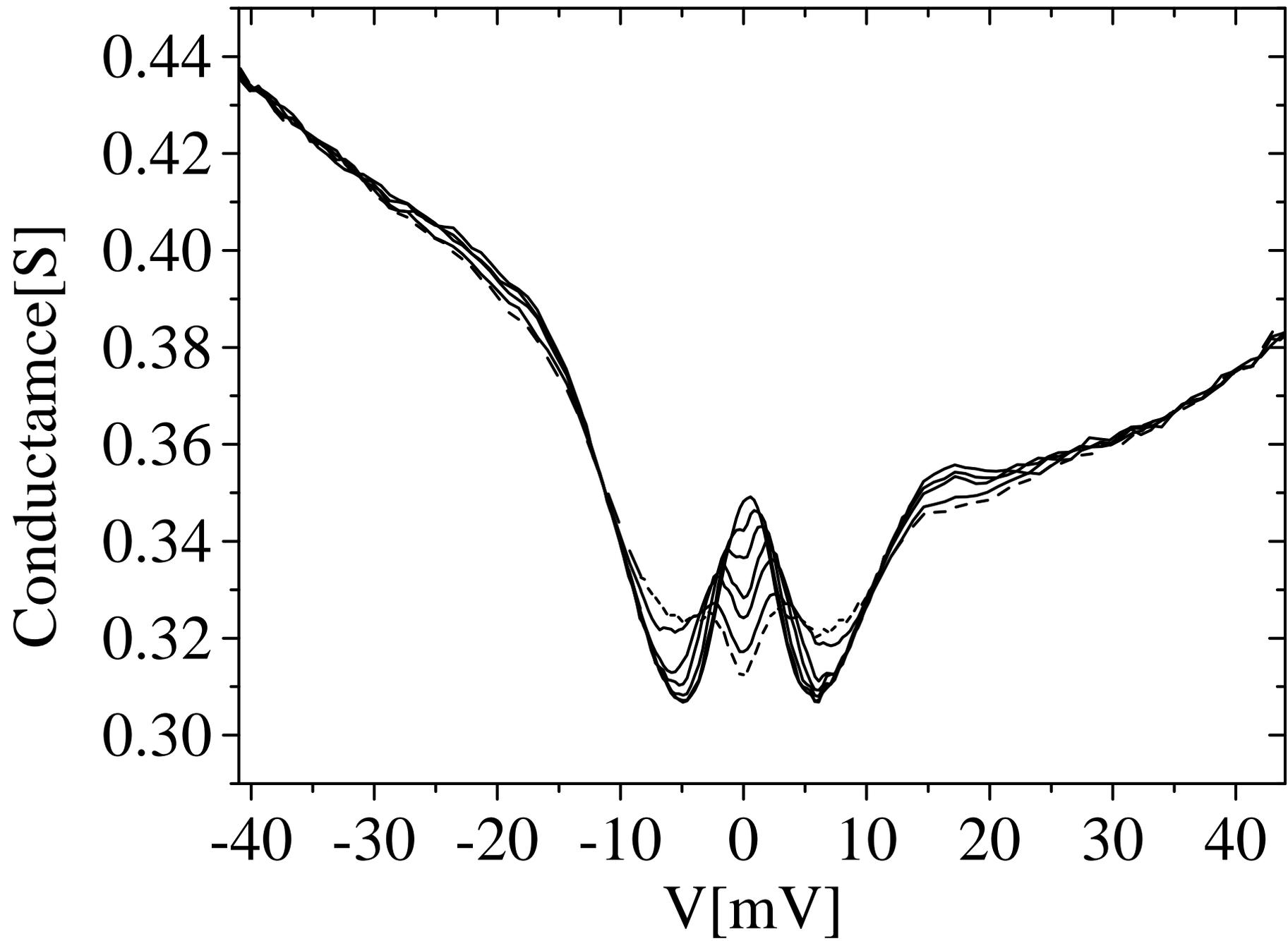

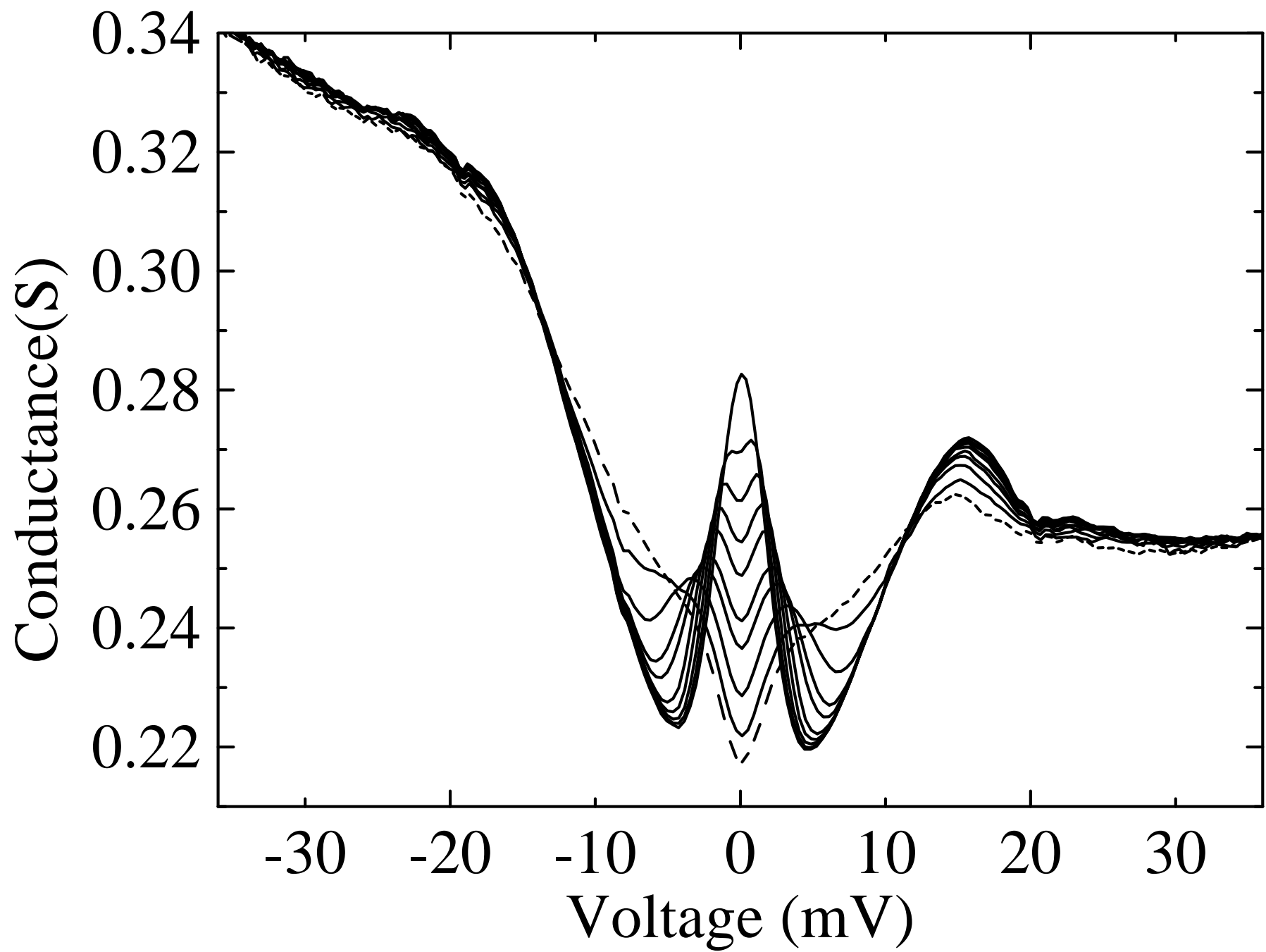

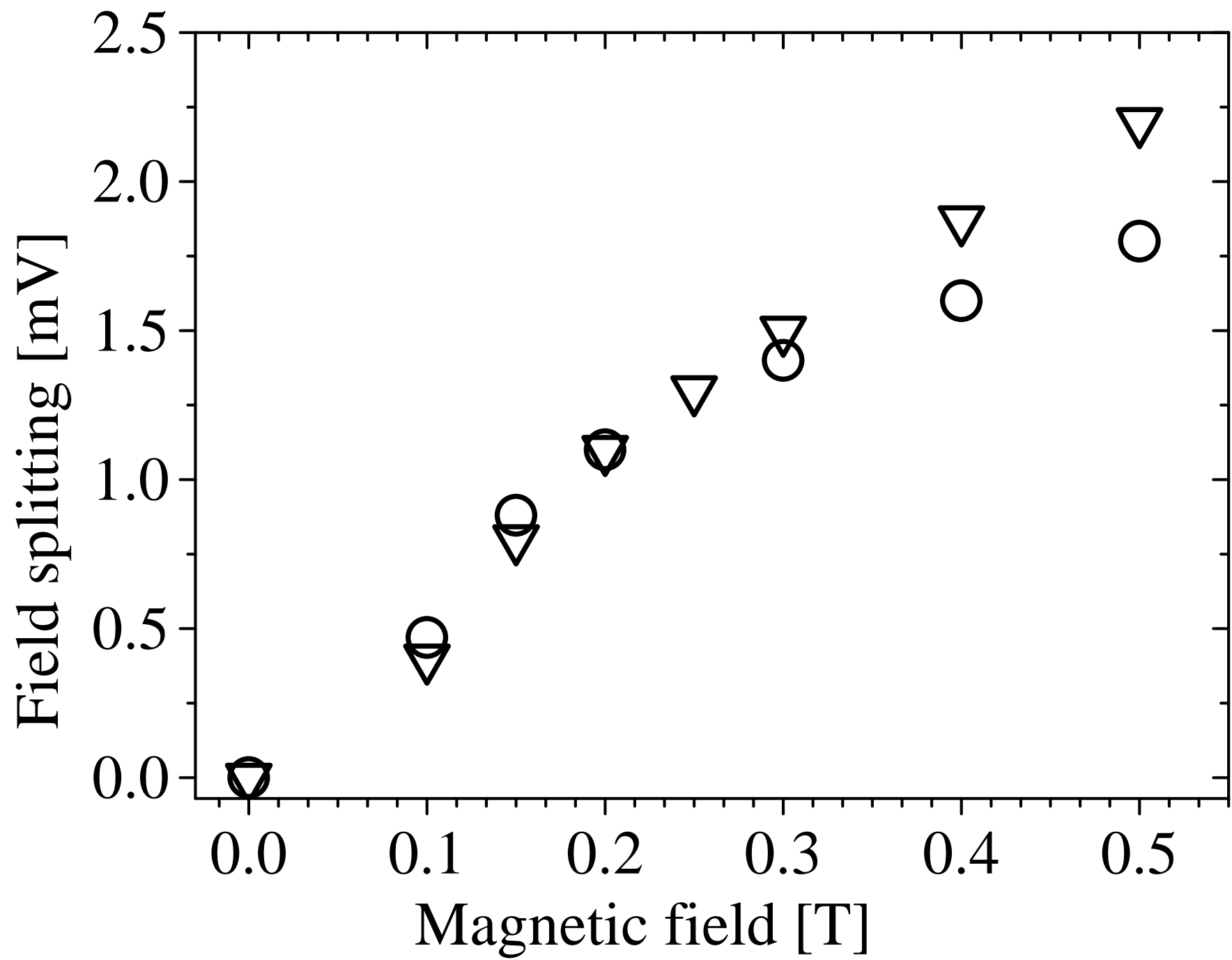

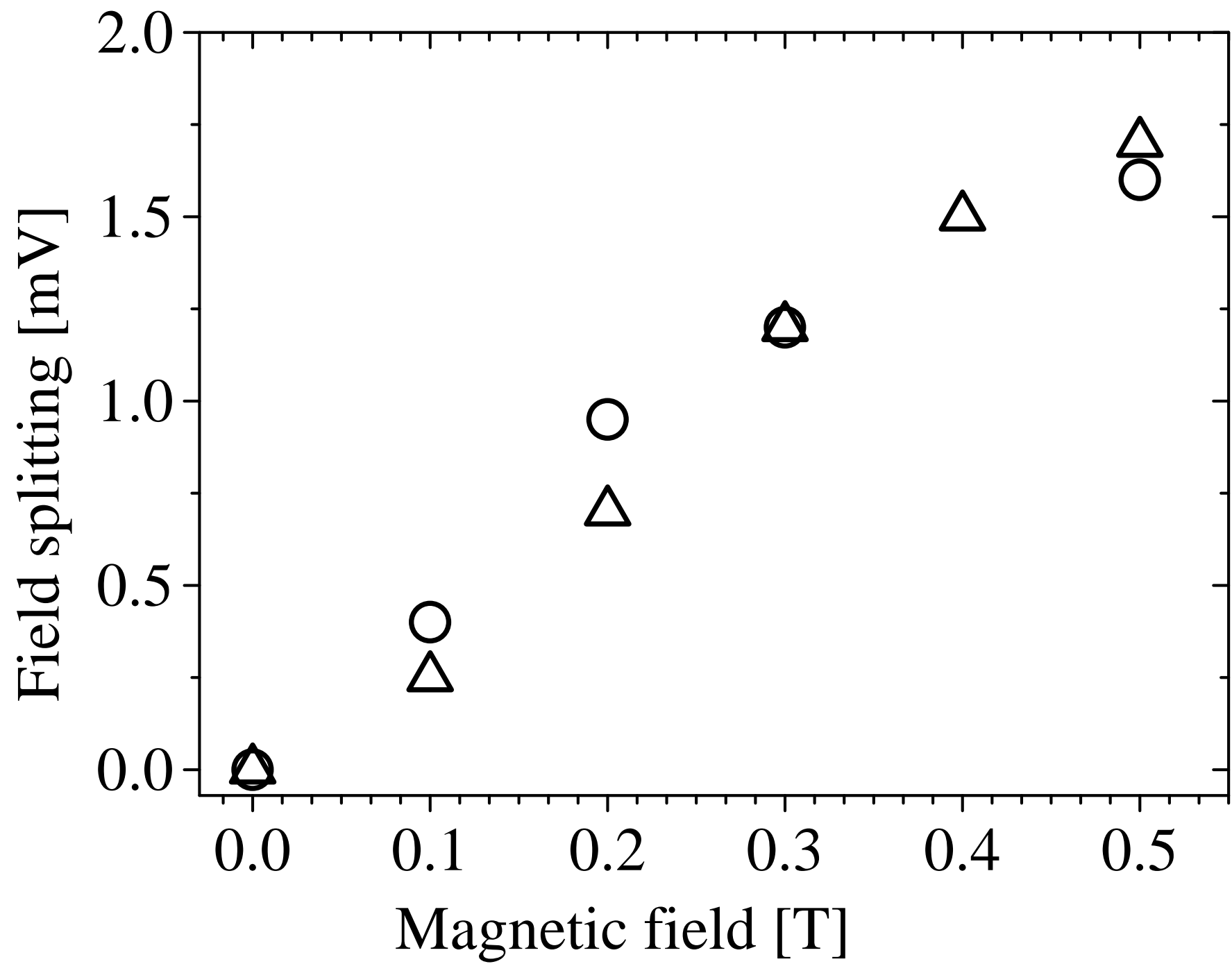

# Thickness independence of field induced time reversal symmetry breaking in $Y_1Ba_2Cu_3O_{7-\delta}$ thin films.

Y. Dagan and G. Deutscher

School of Physics and Astronomy, Raymond and Beverly Sackler Faculty of Exact Sciences, Tel Aviv University, 69978 Tel Aviv, Israel.

**Abstract**

We have measured the field dependence of the Zero Bias Conductance Peak (ZBCP) of tunnel junctions prepared on (110) oriented $Y_1Ba_2Cu_3O_{7-\delta}$ films of various thickness. We have found that the field splitting of the ZBCP, believed to be due to the Time Reversal Symmetry Breaking (TRSB) of Andreev surface bound states, is independent of film thickness. This rules out that Meissner screening currents are at the origin of TRSB. We conclude that TRSB must, instead, be due to a field-induced modification of the Order parameter. Different possibilities involving an imaginary component are discussed.

One of the new properties of the High $T_c$ cuprates, by now well established, is the dominant *d*-wave symmetry character of the order parameter.[1] An interesting consequence of this symmetry is the formation of low energy surface bound states at the boundary of the specimen, at or near the Fermi level. These states result from the interference between quasi-particles, which, upon reflection at the boundary, undergo Andreev reflections from lobes of the OP having opposite signs.[2] Andreev surface bound states increase the conduction of tunnel junctions at small bias (below the gap). The characteristics of the junction then present a ZBCP.[3] This is by now well understood, and well-established experimentally.[4,5,6]

If, for specificity, we consider a surface whose normal is parallel to the direction where the OP has a node, say [110], Andreev surface bound states carry currents along the $[\bar{1}10]$ and $[1\bar{1}0]$ directions. States carrying currents in these opposite directions are degenerate and cancel each other. But, if for any reason, there exists at the surface a net current, this degeneracy is lifted. The energy of half the surface bound states is shifted upwards, the other downwards. The ZBCP is then split, the splitting being proportional to the surface current.

A field induced splitting was observed by several groups [4,5,6,7]. An interpretation of this effect was given in terms of field induced Meissner surface currents.[8] As shown later, this splitting is indeed maximal when the field is applied perpendicular to the $CuO_2$ planes, [5,6] which is consistent with this model since in that orientation Meissner currents flow in the $CuO_2$ planes. At low fields, the splitting is in general proportional to the applied field. In the framework of the above model, the slope of the splitting at low fields is related to the width of the tunneling cone of the specific junction used in the experiment. This cone limits the component of the momentum of tunneling quasi-particles, which is parallel to the surface, to values equal to or smaller than $k_F.\sin\theta$, where θ is the half width of the tunneling cone and $k_F$ is the wave number at the Fermi level. This limits the energy shift to values equal to or smaller than $p_s.v_F\sin\theta$, where $p_s$ is the superfluid momentum of the Meissner currents and $v_F$ is the Fermi velocity. The opening of the tunneling cone can be retrieved from the experiment. From their measured slopes, Covington et al.[4] and Krupke et al.[5] calculate an opening of a few degrees, somewhat smaller than typical openings of tunneling cones (10 to 15 degrees, see Ref. ([9])).

More serious difficulties with this interpretation were noted. By investigating samples having different doping levels, Dagan et al.[10] and Deutscher et al.[11] noted that the field splitting is in fact doping dependent. It is not clear why the opening of the tunneling cone, which is the only free parameter in the theory of Fogelström et al., should vary as a function of doping.

Because of these difficulties, we decided to check directly the Meissner screening model of ref.(8) by varying the film thickness. With the magnetic field applied parallel to the sample surface, the superfluid momentum is reduced when the film thickness is made smaller that the London penetration depth, λ. Its value near the surface is given by:[12]

$$p_s = e\lambda H \tanh\left(\frac{d}{2\lambda}\right) \qquad (1)$$

Where e is the electron charge, d is the film thickness and H is the external magnetic field at the surface of the sample. Hence, the field splitting should be reduced according to the thickness dependence of $p_s$. (In fact, as shown in ref 8, surface currents flow counter to the Meissner currents, but they are proportional to them.)

We have measured the field splitting of the ZBCP in junctions prepared on (110) oriented films of $Y_1Ba_2Cu_3O_{7-\delta}$. Film thickness ranged from 2400Å down to 800Å, respectively larger and smaller than the penetration depth, which we take to be equal to 1800Å as found in thin films.[13] We have compared the field splitting rate between junctions having approximately similar resistance per unit area, hence similar cone openings. We have found no correlation between the splitting rate and the film thickness. Films used for these experiments were all slightly underdoped, and displayed no spontaneous splitting.[11]

(110) oriented $Y_1Ba_2Cu_3O_{7-\delta}$ were grown on (110) $SrTiO_3$ substrates using DC off-axis sputtering with a $Pr_1Ba_2Cu_3O_7$ template used in order to reduce (103) outgrowths as described in [14]. The films were examined using x-ray diffraction, which showed only peaks corresponding to the (110) orientation. We observed an anisotropy in the electrical resistivity, as shown in,[15] from which we conclude that the c - axis is in plane oriented. The films have a critical down-set (zero resistance) temperature, $T_c$=86K – 88.8K, they have been slightly underdoped by annealing in a low oxygen atmosphere. The film thickness was determined in an atomic force microscope measurement of a step created by etching part of the film in $H_3PO_4$ solution. Junction fabrication and characterization and the measurement technique are the same as in (Ref. 15).

Fig.1 shows the characteristics at various magnetic fields, at a temperature T=4.2K, of a junction prepared on a thick film (d=2400Å, $T_c$=88.6K down-set). In Fig.2 we present the characteristics at various magnetic fields at T=4.2K of a junction prepared on a thin film (d=800Å, $T_c$=86K down-set).

Figure 3 shows the splitting of the ZBCP, δ, (measured as the position of the peak at positive biases) as a function of magnetic field for a set of two samples having a thickness of 800Å (triangles $T_c$=88.5K down-set) and 2400Å (circles $T_c$=88.8K). A similar pair of samples with the same thickness values is shown in figure 4, here $T_c$=86K for both samples.

One can see that the splitting rate of the thick films is the same as of the thinner ones at low fields, while according to the thickness values, it should be 2.7 times larger for the thicker ones, if the splitting was controlled by Meissner currents. We have in fact found that the field splitting rate is correlated with the doping level rather than with the film thickness.[10, 11, 16]

We must conclude that the field splitting of the ZBCP is not due to Meissner currents. (The film thickness does have some influence on the high field behavior, but not on the initial slope of the splitting.[17]) The origin of the field splitting must then lie in a field-induced modification of the order parameter. A field induced imaginary component, resulting in surface currents, can be energetically favorable if it involves a magnetic moment. This is the case for an $id_{xy}$ component but not for an $is$ one, as pointed out by Laughlin.[18]

We are indebted to Z. Barkay for helping us with the thickness measurements.
This work was supported in part by the Heinrich Hertz - Minerva Center for High Temperature Superconductivity, by the Israeli Science Foundation and by the Oren Family Chair of Experimental Solid State Physics.

Figure 1.

The conductance versus voltage at 4.2K at various magnetic fields: 0T, 0.1T, 0.2T, 0.5T, 1T, 4T, and 6T (dashed). The film had a zero resistance transition temperature of 88.8K and a thickness of 2400Å.

Figure 2.

The conductance versus voltage at 4.2K at various magnetic fields: 0T, 0.2T, 0.3T, 0.5T, 0.7T, 1T, 2T, 4T, and 6T (dashed). The film had a zero resistance transition temperature of 86K and a thickness of 800Å.

Figure 3.

The position of the split ZBCP at positive biases as a function of magnetic field for a pair of samples having respectively $T_c$=88.8K and $T_c$=88.5K and a different thickness: 2400Å (circles) and 800Å (triangles).

Figure 4.

The position of the split ZBCP at positive biases as a function of magnetic field for a pair of samples having the same $T_c$=86K and different thickness: 2400Å (circles) and 800Å (triangles).